\documentclass
[twocolumn,aps,prc,amsmath,amssymb,floatfix]
{revtex4-2}
\usepackage{CJK}
\usepackage[dvips]{graphicx}         %
\usepackage{bm}                      
\usepackage{mathptmx}                
\usepackage{dcolumn}                 
\usepackage{multirow}                
\usepackage{xcolor}                  %
\usepackage[
breaklinks,pdfstartview=FitH,CJKbookmarks=true,
bookmarksnumbered=true,bookmarksopen=true,pdfborder={0 0 1},
colorlinks=true,linkcolor=blue,urlcolor=blue,anchorcolor=blue,citecolor=blue
            ]{hyperref}

\begin{document}

\def\be{\begin{equation}} \def\ee{\end{equation}}
\def\bal#1\eal{\begin{align}#1\end{align}}
\def\bse#1\ese{\begin{subequations}#1\end{subequations}}
\def\rra{\right\rangle} \def\lla{\left\langle}
\def\rv{\bm{r}} \def\tv{\bm{\tau}} \def\sv{\bm{\sigma}}
\def\tt{(\tv_1\cdot\tv_2)} \def\ss{(\sv_1\cdot\sv_2)}
\def\non{\nonumber}
\def\ra{\rightarrow}
\def\arsinh{\mathop{\text{arsinh}}}
\def\al{\alpha}
\def\la{\Lambda}
\def\eps{\epsilon}
\def\ms{\,M_\odot}
\def\mmax{M_\text{max}}
\def\l14{\Lambda_{1.4}}
\def\r14{R_{1.4}}
\def\r14{R^{(N)}_{1.4}}
\def\lr14{\l14 / \la_\text{fit}(\r14)}
\def\fm3{\;\text{fm}^{-3}}
\def\km{\;\text{km}}
\def\mev{\;\text{MeV}}
\def\gev{\;\text{GeV}}
\def\mfm{\text{MeVfm}^{-3}}
\def\fb#1{\textcolor{blue}{#1}}
\def\hcd#1{\textcolor{magenta}{#1}}
\long\def\hj#1{\color{red}#1\color{black}}
\def\OFF#1{}

\title{
Dark matter effects on the properties of neutron stars:
compactness and tidal deformability
}

\begin{CJK*}{UTF8}{gbsn}

\author{Hong-Ming Liu (刘宏铭)$^{1,2}$} 
\author{Jin-Biao Wei (魏金标)$^3$}
\author{\hbox{Zeng-Hua Li (李增花)$^{1,2}$}} \email[]{zhli09@fudan.edu.cn}

\affiliation{
$^1$\hbox{
Institute of Modern Physics, Key Laboratory of Nuclear Physics
and Ion-Beam Application, MOE, Fudan University,}\\
{Shanghai 200433, People's Republic of China}\\
$^2$\hbox{
Shanghai Research Center for Theoretical Nuclear Physics,
NSFC and Fudan University, Shanghai 200438, China}\\
$^3$\hbox{
School of Mathematics and Physics, China University of Geosciences,
Lumo Road 388, 430074 Wuhan, China}
}

\author{G. F. Burgio$^4$}
\author{H. C. Das$^4$}
\author{H.-J. Schulze$^4$}
\affiliation{
$^4$\hbox{
Istituto Nazionale di Fisica Nucleare, Sezione di Catania,
Dipartimento di Fisica, Universit\'a di Catania,}\\
{Via Santa Sofia 64, 95123 Catania, Italy}
}

\date{\today}

\begin{abstract}
We systematically study the observable properties of
dark-matter admixed neutron stars,
employing a realistic nuclear EOS in combination with self-interacting fermionic
dark matter respecting constraints on the self-interaction cross section.
Deviations from universal relations valid for nucleonic neutron stars
are analyzed over the whole parameter space of the model
and unequivocal signals for the presence of dark matter in neutron stars
are identified.
\end{abstract}

\maketitle
\end{CJK*}

\section{Introduction}

Dark matter (DM),
realized by yet unknown elementary particles within or beyond the standard model
\cite{Zwicky09,Bertone05,Feng10},
is one of the most enigmatic aspects of current astrophysics
\cite{Trimble87,Bergstrom00}.
It must make up nearly 90\% of the matter in the Universe
in order to explain observations at galactic and super-galactic scales
\cite{Begeman91,Abdalla09,Abdalla10,Wittman00,Massey10}.
Most of its properties like its mass and
interactions with other particles are presently unknown
\cite{Henriques90,Ciarcelluti11,Guever14,Raj18}.

Theoretically several kinds of bosonic and fermionic DM candidates
have been hypothesized,
such as a weakly-interacting massive particle (WIMP)
\cite{Goldman89,Andreas08,Kouvaris08,Kouvaris12,Bhat20},
especially neutralino  
\cite{Hooper04,Panotopoulos17,ADas19,
Das20,Das21b,Das21c,Das22,Kumar22,Lourenco22},
asymmetric dark matter (ADM)
\cite{Kouvaris11,McDermott12,Gresham19,Ivanytskyi20}
like mirror matter
\cite{Okun07,Sandin09,Hippert22},
axion
\cite{Duffy09,Balatsky22}, and
strangelets
\cite{Jacobs15,Ge19,Vandevender21,Zhitnitsky21}.

A particular environment to reveal features of DM are neutron stars (NSs),
the densest objects known in the Universe,
for which more and more precise observational data become available.
Their hypothetical capability to accumulate DM might provide possibilities
to deduce related DM properties.
It is thus of great interest to theoretically analyze
DM-admixed NS (DNS) models,
combining a more or less well-known nuclear-matter (NM) EOS
with a hypothetical DM EOS
within a general-relativistic two-fluid approach
\cite{Kodama72,Comer99,Sandin09}.

The fundamental theoretical challenge of the eventual presence of DM in NSs
is the fact that their mass-radius relation is not anymore a unique function,
but depends on an additional degree of freedom, the DM content.
The absence of any knowledge regarding the nature of DM,
combined with the persisting uncertainty of the high-density NM EOS,
then renders any theoretical conclusion regarding either DM or high-density NM
doubtful.
The simplest example would be the observation of a very massive `NS',
which could simply be caused by the presence of a massive DM halo,
but also by a very stiff NM EOS.
Therefore theoretical methods have to be devised
to unequivocally identify the presence and quantity of DM in NSs.

It is generally believed that most or all observed 'NSs' are such,
namely standard hadronic stars with very little admixture of DM.
We note at this point that current estimates of the acquired DM content
by accretion during the NS lifetime yield extremely small results of
$\lesssim10^{-10}\ms$
\cite{Goldman89,Kouvaris08,Kouvaris10,Kouvaris11,McDermott12,Guever14,
Bramante15,Baryakhtar17,Deliyergiyev19,Ivanytskyi20},
which would be unobservable.
The existence of DNSs with large DM fractions
(of the order of percent or larger)
assumed in this article,
therefore requires exotic capture or formation mechanisms
\cite{Ciarcelluti11,Goldman13,Kouvaris15,Eby16,Maselli17,Ellis18,
Nelson19,Deliyergiyev19,DiGiovanni20},
which remain so far very speculative.
Keeping this in mind,
we study in this work in a qualitative manner
DNSs with arbitrary DM fraction up to 100\%,
corresponding to pure dark stars (DSs)
\cite{Colpi86,Schunck03,Liebling12,Eby16}.

The effects of DM on the properties of NM and NSs
have been intensely investigated in recent years.
Apart from its fundamental impact on NS mass and radius
\cite{Henriques90,Leung11,Li12,Tolos15,Delpopolo20b,Das21,Yang21,Kain21},
a large number of works studied the possibility of DM acquisition by NSs
\cite{Lopes11,Delpopolo20,Bell21,Maity21,Anzuini21,Bose22}
and related phenomena like heating
\cite{Kouvaris08,Kouvaris10,Gonzales10,Bertoni13,Baryakhtar17,Raj18,
Garani21,Coffey22,Fujiwara22}
or internal black hole formation and collapse
\cite{Goldman89,Sandin09,Lavallaz10,Kouvaris12,McDermott12,
Bramante13,Bramante14,Bramante15,Ivanytskyi20,Liang23}.
Recently, more quantitative studies have been performed,
like the DM effects on the derived properties of NM
\cite{Das20},
or the validity of universal relations between the
NS compactness $M/R$ and the
moment of inertia $I$,
tidal deformability $\la$,
and quadrupole moment $Q$ 
\cite{Yagi13,Yagi13b,Yagi17,Godzieba21},
in the presence of DM
\cite{Maselli17,Routaray23,Thakur23}.
Refs.~\cite{Zhang20,Zhang22}
examined the possibility of the LIGO/Virgo events
GW170817 \cite{Abbott17} and GW190425 \cite{Abbott20}
being realized by a DNS scenario.
Many recent works
\cite{Maselli17,Ellis18,ADas19,Nelson19,Quddus20,Husain21,Das21b,Das21c,
Das22,Leung22,Lourenco22,Karkevandi22,Dengler22,Hippert22,Collier22,Dutra22,
Diedrichs23,Karkevandi23}
focused on DM effects on the NS tidal deformability
and related observables \cite{Das21d,Emma22,Bauswein23,Zhang23},
which are directly accessible by recent and future GW detectors.
The impact of DM on the pulsar x-ray profile \cite{Miao22,Shakeri24}
and on NS cooling processes \cite{Bhat20,Kumar22}
were also examined recently.

The purpose of the present article
is to continue and extend our previous study
of DNS optical radii and tidal deformability \cite{Liu23}
over the full parameter space of a given DM model.
We will now assume self-interacting fermionic DM particles
with an interaction strength constrained by observational limits
on the self-interaction cross section.

The optical radius $R_N$ is perhaps the most relevant
and at the same time the most easily accessible observable of a NS
(apart from the gravitational mass $M$),
and therefore merits particular attention before studying more intricate
features of a NS.
We will also analyze the validity and breakdown of universal relations
between stellar compactness $M/R_N$ and tidal deformability $\Lambda$.
The purpose is to devise unequivocal methods to deduce and quantify the
presence of DM in NSs.

This article is organized as follows.
In the next section~\ref{Sec.2}, the EOSs
of ordinary NM and DM used in this work are briefly described.
The detailed calculations and discussion are presented in Sec.~\ref{Sec.3}.
A summary is given in Sec.~\ref{Sec.4}.

\section{Formalism}
\label{Sec.2}

\subsection{Equation of state for nuclear matter}

As in \cite{Liu23},
we employ here the latest version of a Brueckner-Hartree-Fock (BHF) EOS
obtained with the Argonne V18 $NN$ potential and compatible three-body forces
\cite{Li08a,Li08b,Liu22},
see \cite{Baldo99,Baldo12} for a more detailed account.
This EOS is compatible with all current low-density constraints
\cite{Wei20,Burgio21,Burgio21b}
and in particular also with those imposed on NS maximum mass
$\mmax>2\ms$
\cite{,Antoniades13,Arzoumanian18,Cromartie20},
radius $R_{1.4}\approx11-13\,$km
\cite{Riley21,Miller21,Pang21,Raaijmakers21},
and tidal deformability $\Lambda_{1.4}\approx70-580$
\cite{Abbott17,Abbott18,Burgio18,Wei19}.

Therefore observation of compact objects violating these constraints
will be interpreted as indicating DM admixture
according to the following detailed analysis.

\begin{figure}[t]
\vskip-12mm
\centerline{\hskip4mm\includegraphics[scale=0.55]{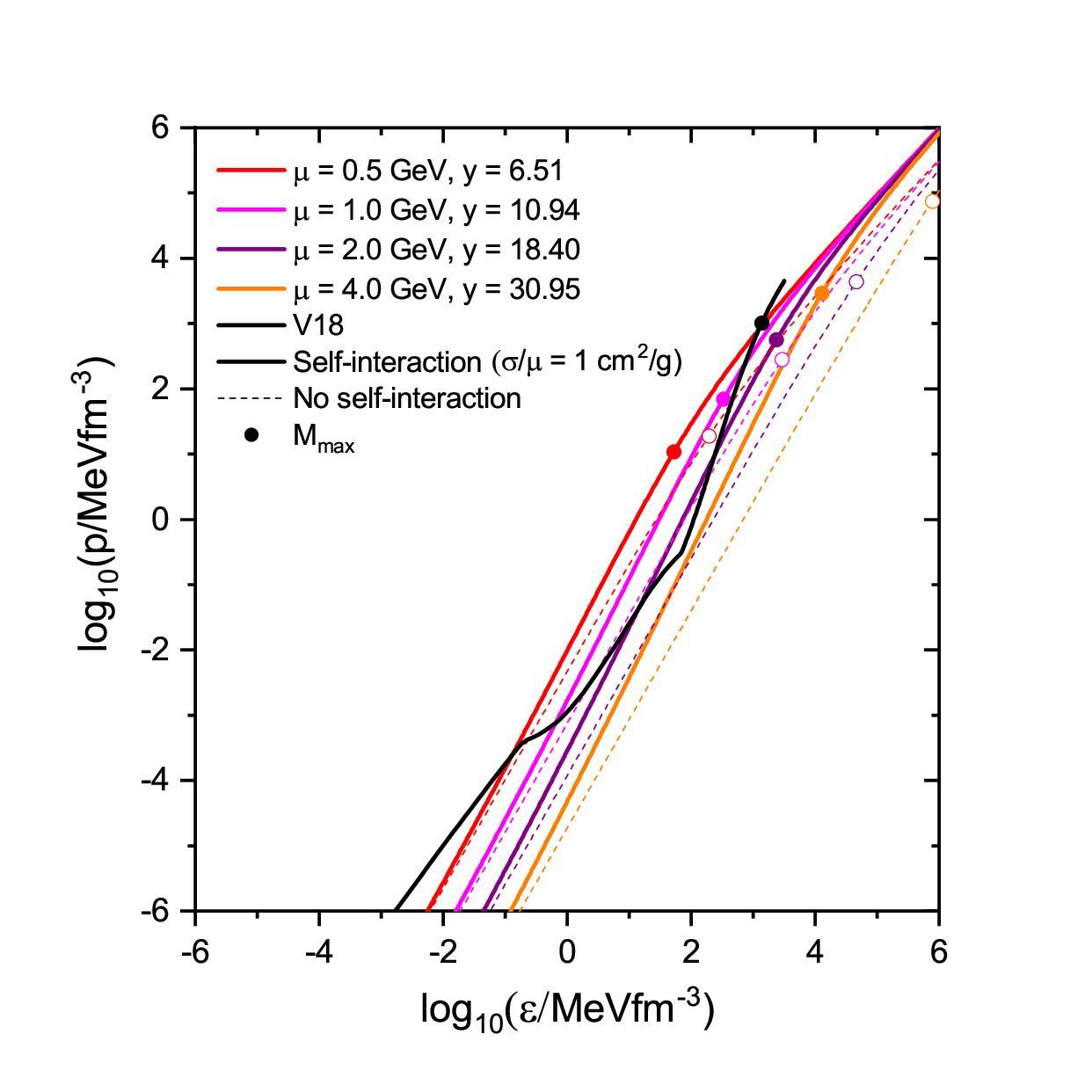}}
\vskip-9mm
\caption{
EOSs of different pure DM models
and the nuclear V18 EOS.
The markers indicate the values of the maximum-mass $\mmax$ configurations.
The interaction parameter $y$, Eq.~(\ref{e:ymu}), is also listed.
\hj{}
}
\label{f:eos}
\end{figure}

\begin{figure}[t]
\vskip-12mm
\centerline{\hskip1mm\includegraphics[scale=0.46]{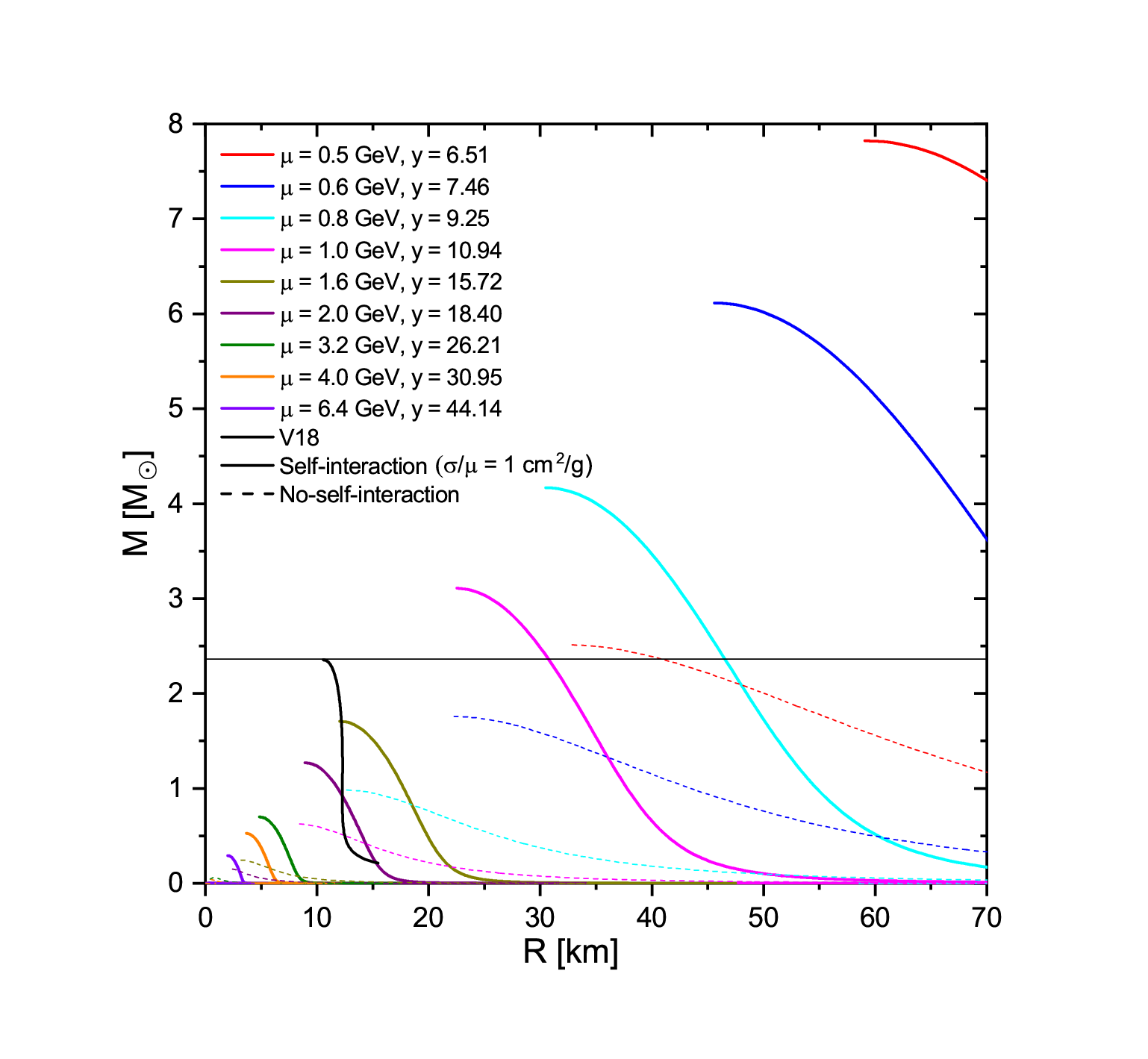}}
\vskip-10mm
\caption{
Mass-radius relations of pure dark stars
for different values of the DM particle mass $\mu$,
with (solid) and without (dashed) self-interaction,
in comparison with the NM EOS V18 (solid black).
The horizontal black line indicates the maximum mass of a pure NS,
$\mmax=2.36\ms$.
}
\label{f:mr}
\end{figure}

\begin{figure*}[t]
\vskip-13mm
\centerline{\hskip-1mm\includegraphics[scale=0.46]{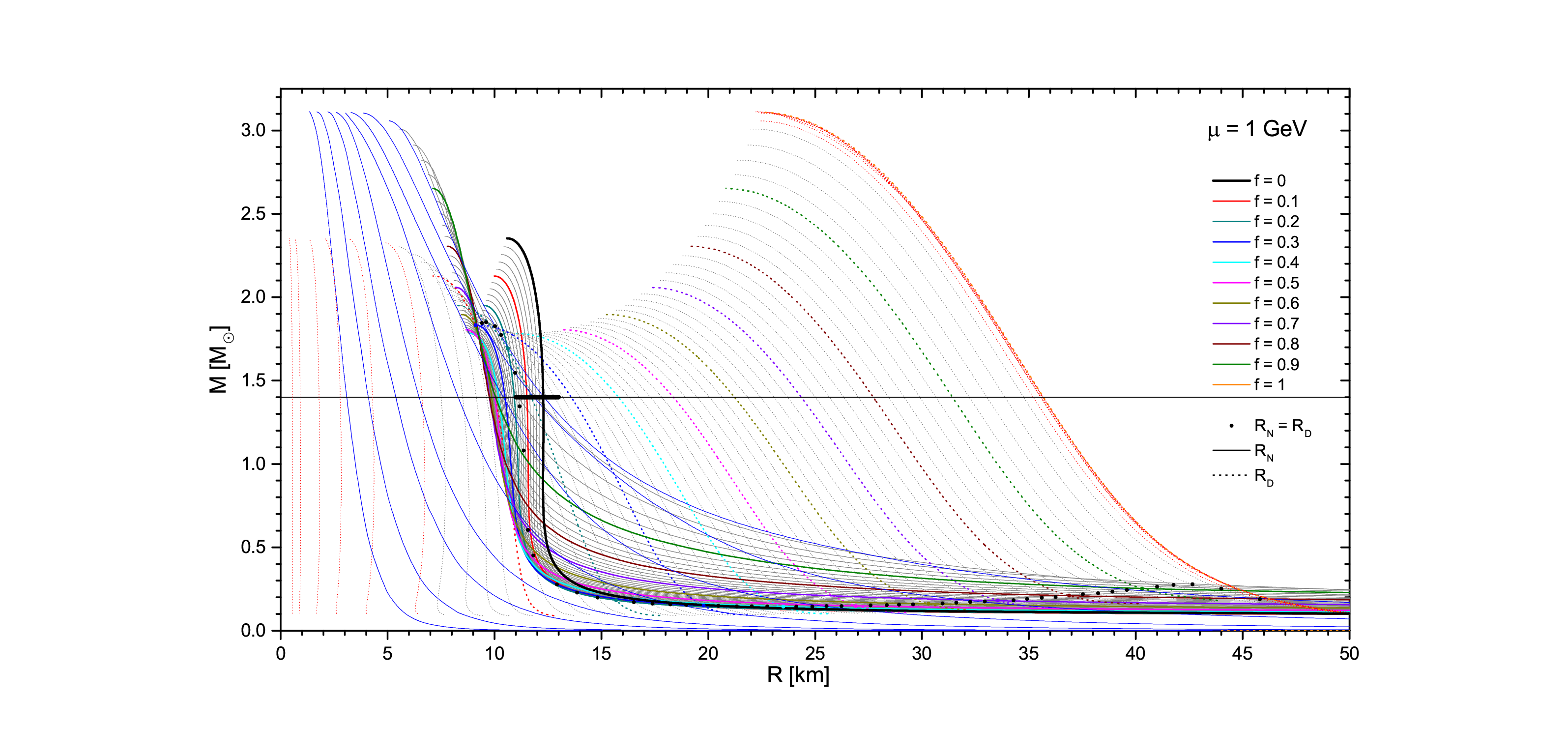}}
\vskip-14mm
\caption{
Total gravitational mass $M$ as function of nuclear (solid curves)
or dark (dotted curves) radius
for the $\mu=1\gev$ model
with different DM fractions $f=M_D/M$.
Markers indicate $R_D=R_N$ configurations.
The range of NS radii $R_{1.4}=11$--$13$ km is represented by a horizontal bar
on the $M=1.4\ms$ line.
The values of $f$ are in intervals of 0.02,
apart from those close to the boundaries:
$f=10^{-2,-3,\ldots,-7}$
(thin dotted red curves)
and
$f=1-10^{-2,-3,\ldots,-9}$
(thin solid blue curves).
}
\label{f:mrnd}
\end{figure*}

\OFF{
\begin{figure*}[t]
\vskip-8mm
\centerline{\hskip-10mm\includegraphics[scale=0.33]{dmf34}}
\vskip-9mm
\caption{
Mass-radius relations as a function of DM fraction $f=M_D/M$
for four values of the DM particle mass $\mu=0.8,1.6,3.2,6.4\gev$.
Total gravitational mass $M$ (solid curves)
and baryonic mass $M_B$ (dotted curves)
are shown as functions of the outer radius $R=\max(R_N,R_D)$.
Vertical lines are to guide the eye.
Horizontal lines indicate $M=1.4$$M_\odot$ and $M=M_\text{max}=2.36\ms$
for the pure NS.
Markers indicate the $R_D=R_N$ stars.
Note the different $R$ scales.
}
\label{f:mrf}
\end{figure*}
}

\begin{figure*}[t]
\vskip-13mm
\centerline{\hskip-25mm\includegraphics[scale=0.34]{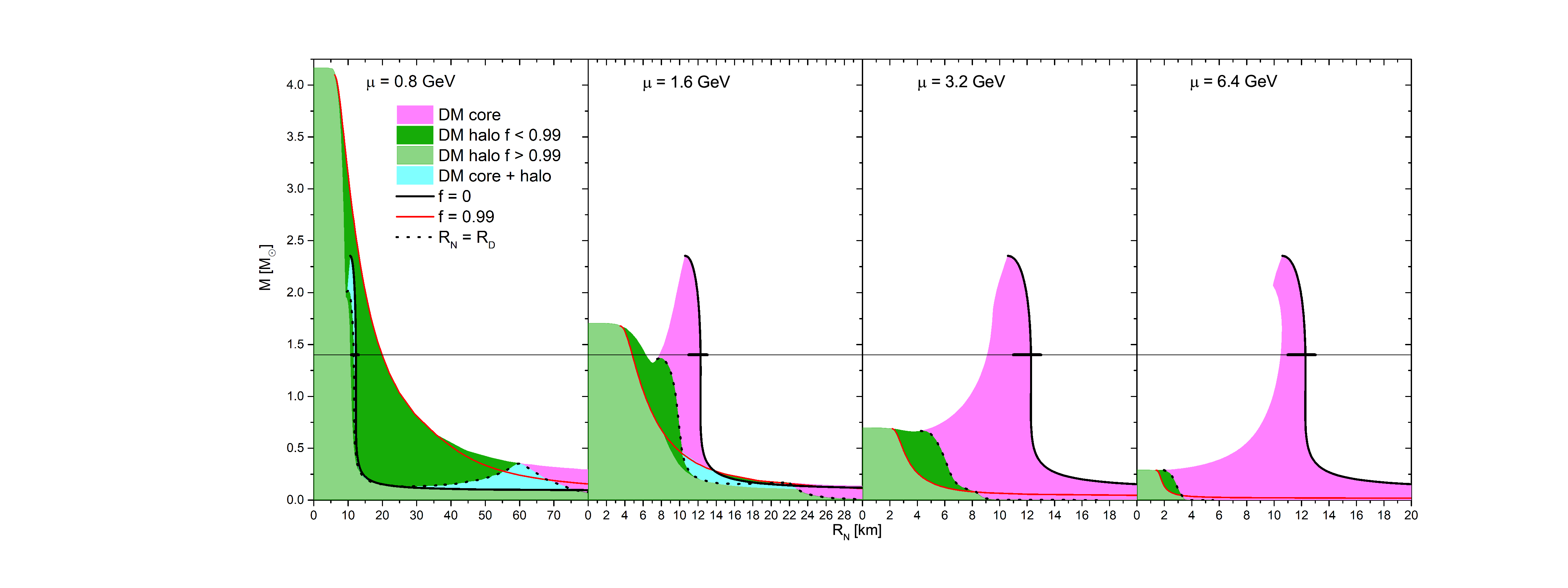}}
\vskip-12mm
\caption{
The domains of stable DM-core (magenta shading),
DM-halo (green),
and both (cyan)
DNSs in the $(M,R_N)$ plane
for different DM models.
Note the different $R_N$ axes.
See extended discussion in the text.
}
\label{f:dns}
\end{figure*}

\begin{figure}[t]
\vskip-17mm\centerline{\hskip-0mm\includegraphics[scale=0.35]{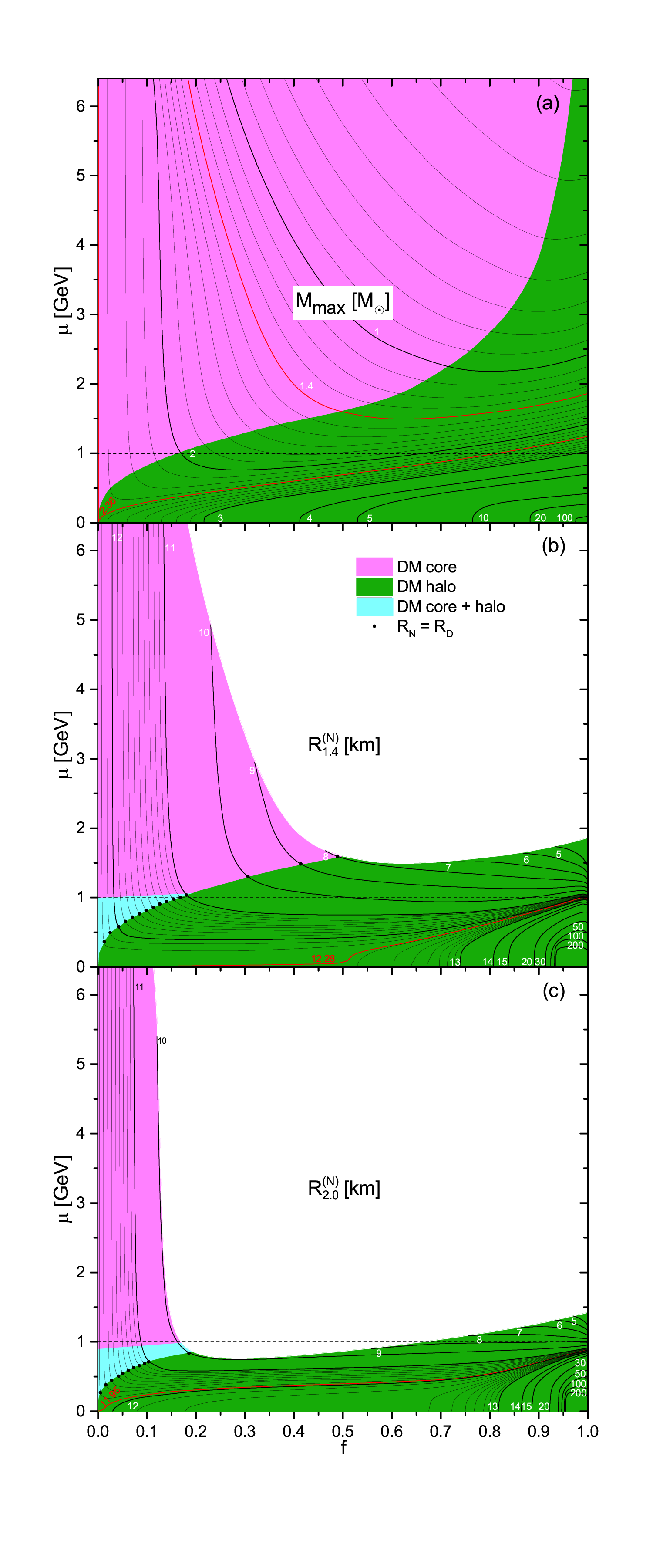}}\vskip-17mm
\caption{
Contour plots of
(a) $\mmax$
(thin contours are in intervals of $0.1\ms$;
2.36 is the value for the pure NS)
and (b,c) optical $\r14$ and $R^{(N)}_{2.0}$
(indicated by numbers in km;
thin contours are in intervals of 0.1 km;
12.28 and 11.95 are the values
for the pure NS),
as functions of $(f,\mu)$.
The color scheme is as in Fig.~\ref{f:dns}.
The horizontal dashed line at $\mu=1\gev$ is to guide the eye.
}
\vskip-6mm
\label{f:r14}
\label{f:mmx}
\end{figure}

\begin{figure}[t]
\vskip-17mm\centerline{\hskip-0mm\includegraphics[scale=0.35]{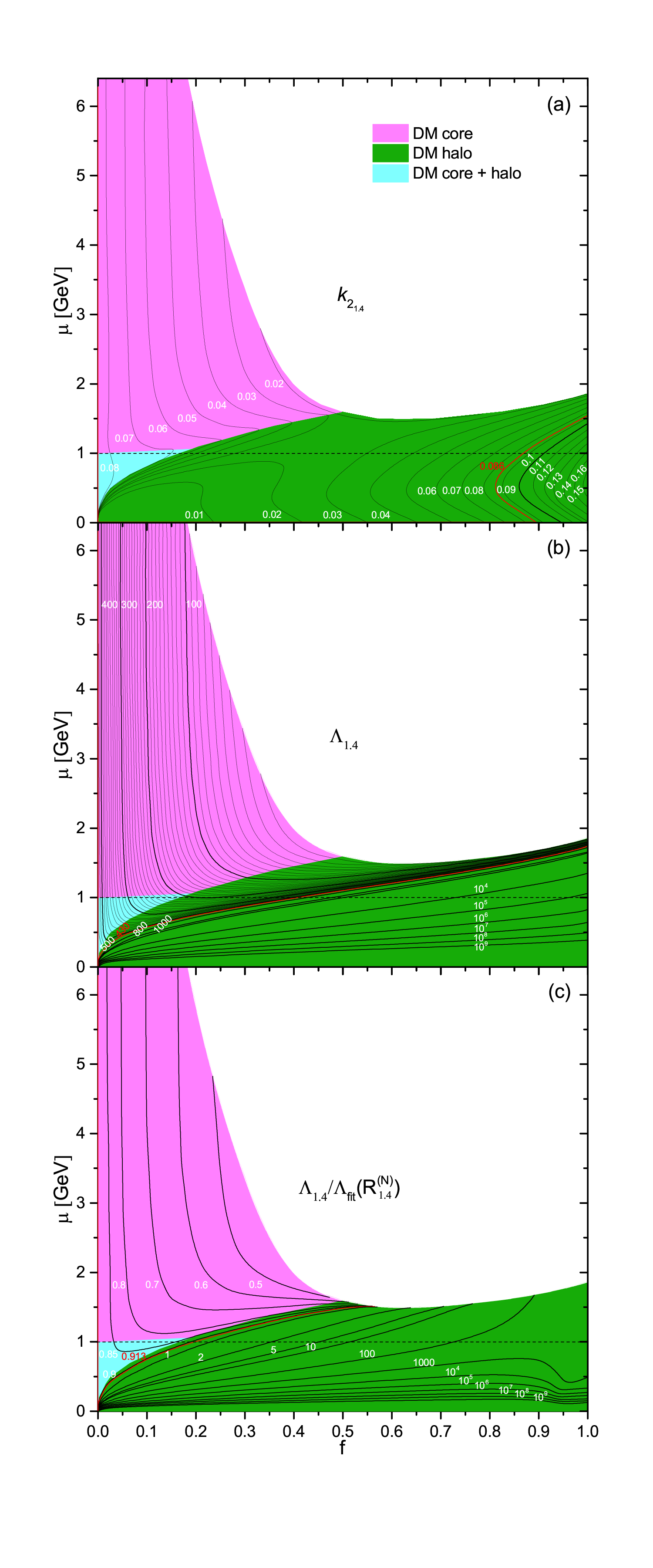}}\vskip-17mm
\caption{
Contour plots of
(a) Love number $(k_2)_{1.4}$
(contours are in intervals of 0.01;
0.086 is the value for the pure NS),
(b) tidal deformability $\l14$
(thin contours are in intervals of 10;
430 is the value for the pure NS),
and
(c) the ratio $\lr14$,
Eq.~(\ref{e:univ2}),
as functions of $(f,\mu)$.
The color scheme is as in Fig.~\ref{f:dns}.
The horizontal dotted line at $\mu=1\gev$ is to guide the eye.
The observation of GW170817 was reported as $\l14=70$--580 \cite{Abbott18}.
}
\vskip-6mm
\label{f:l14}
\end{figure}

\begin{figure*}[t]
\vskip-10mm
\centerline{\hskip8mm\includegraphics[scale=0.6]{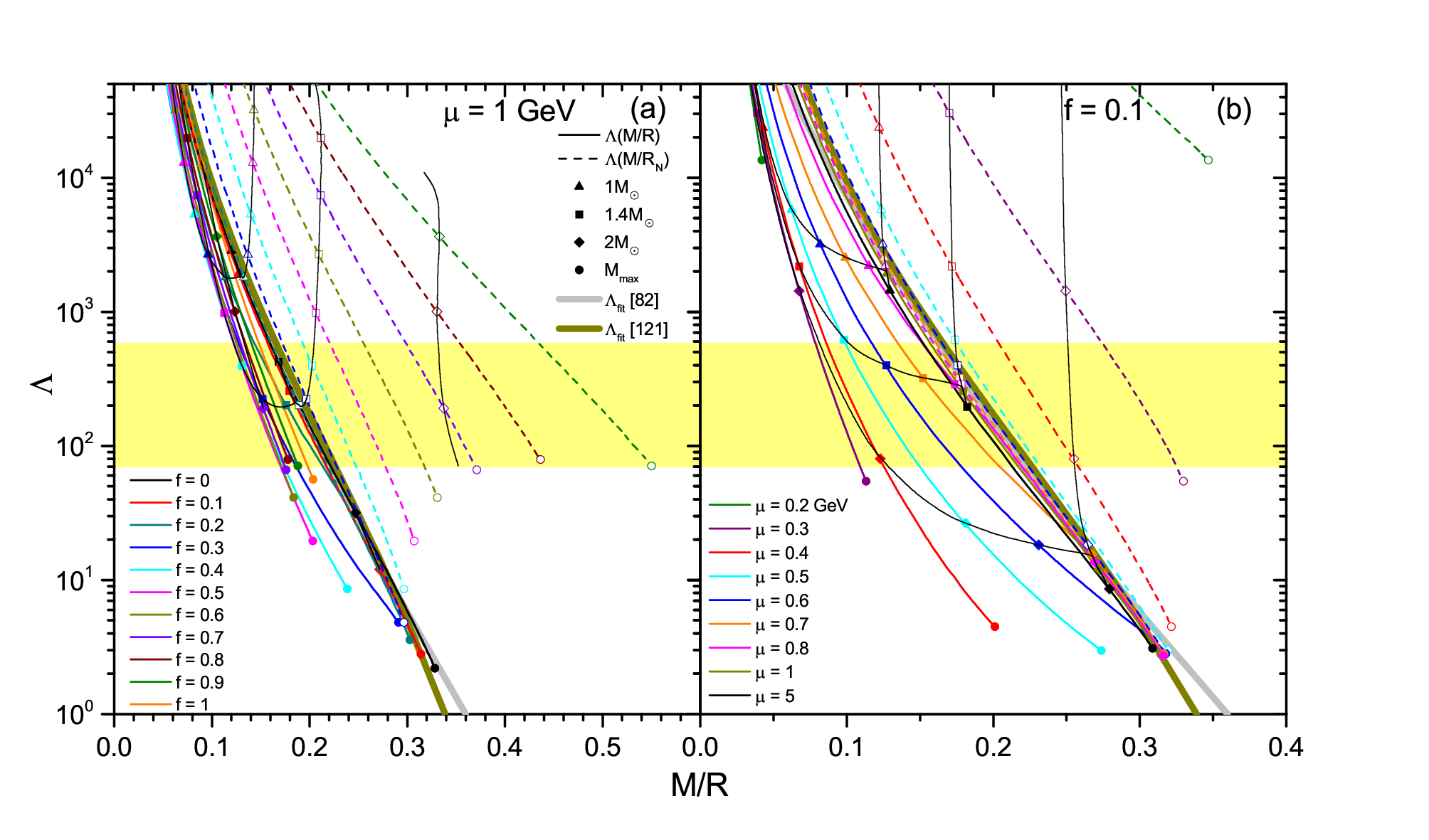}}
\vskip-9mm
\caption{
$\Lambda$ vs.~$M/R$ (dashed curves) and $M/R_N$ (solid curves)
for (a) $\mu=1\gev$, varying $f$
and (b) $f=0.1$, varying $\mu$.
Configurations of fixed $M=1,1.4,2\ms,\mmax$
are indicated by markers and joined by black lines.
The universal relations for pure NSs of \cite{Yagi17,Godzieba21}
are shown as wide curves.
The shaded band indicates the range 70--580 for $M=1.4\ms$
(square markers)
deduced from GW170817 \cite{Abbott18}.
\hj{}
}
\label{f:lmr}
\end{figure*}

\subsection{Equation of state for dark matter}

We employ in this work the frequently used
\cite{Narain06,Li12,Tulin13,Kouvaris15,Maselli17,Ellis18,Nelson19,
Delpopolo20b,Husain21,Leung22,Collier22,Cassing23}
DM model of fermions with mass $\mu$
self-interacting via a repulsive Yukawa potential
\bal
 V(r) &= \al \frac{e^{-m r}}{r} \:
\eal
with coupling constant $\al$ and mediator mass $m$.
Following Ref.~\cite{Narain06} we write
pressure and energy density
of the resulting DM EOS as
\bal
 p_D    &= \frac{\mu^4}{8\pi^2}
 \Big[ x\sqrt{1+x^2}(2x^2\!/3-1) + \arsinh{(x)} \Big]
 + \delta \:,
\\
 \eps_D &= \frac{\mu^4}{8\pi^2}
 \Big[ x\sqrt{1+x^2}(2x^2+1) - \arsinh{(x)} \Big]
 + \delta \:,
\eal
where
\be
 x = \frac{k_F}{\mu} = \frac{(3\pi^2n)^{1/3}}{\mu}
\ee
is the dimensionless kinetic parameter
with the DM particle density $n$,
and the self-interaction term is written as
\bal
 \delta =
 \frac{2}{9\pi^3} \frac{\al \mu^6}{m^2} x^6 &\equiv
  \mu^4 \Big(\frac{y}{3\pi^2}\Big)^2 x^6
 = \Big(\frac{yn}{\mu}\Big)^2 \:,
\label{e:y}
\eal
introducing the interaction parameter
$y^2 = 2\pi\al\mu^2\!/m^2$.
Ref.~\cite{Narain06} contains interesting scaling relations
regarding the EOS and mass-radius relations of pure fermionic DM stars.

Within this model,
$y$ is not a free parameter,
but constrained by observational limits
imposed on the DM self-interaction cross section $\sigma$
\cite{Markevitch04,Kaplinghat16,Sagunski21,Loeb22},
\be
 \sigma/\mu \sim 0.1-10 \;\text{cm}^2\!/\text{g} \:.
\ee
In \cite{Tulin13,Kouvaris15,Maselli17}
it has been shown that the Born approximation
\bal
 \sigma_\text{Born} &= \frac{4\pi\al^2}{m^4}\mu^2 = \frac{y^4}{\pi\mu^2}
\eal
is very accurate for $\mu\lesssim1\gev$
and in any case remains valid in the limit $\al\ra0$ for larger masses.
We therefore employ here this approximation,
choosing for simplicity the fixed constraint
\be
 \sigma/\mu = 1 \;\text{cm}^2\!/\text{g} = 4560/\gev^3 \:,
\ee
which appears compatible with all current observations.
This implies
\bal
 y^4 &= \pi \mu^3 \sigma/\mu \sim \pi (16.58\mu_1)^3
\:,\\
 y &\sim 10.94 \mu_1^{3/4}
\label{e:ymu}
\eal
with $\mu_1\equiv\mu/1\gev$.
After this, the DM EOS depends only on the parameter $\mu$.

This is demonstrated in Fig.~\ref{f:eos},
which compares the DM EOS with and without self interaction for four
typical masses $\mu=0.5,1,2,4 \gev$,
in comparison with the nucleonic EOS.
Markers indicate the configurations of maximum stellar mass.
It can be seen that the self-interaction is always important in the
relevant density domains
and stiffens substantially the EOS.
The nucleonic EOS exhibits
different trends for core, inner crust, and outer crust of a NS.

\subsection{Hydrostatic configuration}

The stable configurations of the DNSs are obtained from a
two-fluid version of the TOV equations
\cite{Kodama72,Comer99,Sandin09}:
\bal
 \frac{dp_D}{dr} &= -[p_D + \eps_D]\frac{d\nu}{dr} \:,
\\
 \frac{dp_N}{dr} &= -[p_N + \eps_N]\frac{d\nu}{dr} \:,
\label{e:tovn}
\\
 \frac{dm}{dr}   &= 4\pi r^2 \eps \:,
\label{e:tovm}
\\
 \frac{d\nu}{dr} &= \frac{m + 4\pi r^3p}{r(r - 2m)} \:,
\label{e:tovnu}
\eal
where $r$ is the radial coordinate from the center of the star, and
$p=p_N+p_D$,
$\eps=\eps_N+\eps_D$,
$m=m_N+m_D$
are the total pressure, energy density, and enclosed mass, respectively.

The total gravitational mass of the DNS is
\be
 M = m_N(R_N) + m_D(R_D) \:,
 \label{eq:gravmass}
\ee
where the stellar radii $R_N$ and $R_D$
are defined by the vanishing of the respective pressures.
There are thus in general two scenarios:
DM-core ($R_D<R_N$) or DM-halo ($R_D>R_N$) stars.

We first analyze the mass-radius relations of pure DSs
in Fig.~\ref{f:mr},
comparing the results with and without self interaction for various
values of $\mu$.
Consistent with the DM EOS,
one notes that the self interaction is always very important and leads to stars
with much larger masses and radii for the same $\mu$.
In this work we only employ values of $\mu\sim{\cal O}(1\gev)$,
such that the corresponding values of $\mmax$ are of the same order
as those of ordinary NSs, $\sim{\cal O}(1\ms)$,
for which the mass-radius relation of the nucleonic V18 EOS
is shown for comparison.
Otherwise the observable effects on NS structure will be very small
\cite{,Liu23}.
We note that in \cite{Narain06} the following scaling relations
for $\mmax$ and the corresponding radius $R(\mmax)$
were derived for $y\gg1$,
\bal
 \mmax/\!\ms &= (0.627+0.269y) /\mu_1^2    
\\
           &= (0.627+2.943\mu_1^{3/4}) /\mu_1^2 \:,
\\
 R(\mmax)/\text{km} &= (8.114 + 1.921y) /\mu_1^2 \:,
\\
                    &= (8.114 + 21.01\mu_1^{3/4}) /\mu_1^2 \:,
\eal
which are reasonably well fulfilled for our range of $\mu$
and associated $y$, Eq.~(\ref{e:ymu}),
given also in the figure.

After analyzing our models of pure NSs and pure DM stars,
we now examine in detail the properties of DNSs
within the two-fluid scenario,
assuming that the two fluids interact only via gravity.
The resulting scenario is very rich
because the stellar configurations depend on the DM fraction $f=M_D/M$.
The following Fig.~\ref{f:mrnd}
gives a detailed account of the possible mass-radius relations of DNSs
(including DM self-interaction)
with $\mu=1\gev$ over the full range of DM fraction $f$.
Both $M(R_N)$ and $M(R_D)$ relations, Eq.~(\ref{eq:gravmass}),
are shown as solid or dotted curves, respectively.

The maximum gravitational mass is $2.36\ms$ for the \hbox{$f=0$} pure NS
(black solid curve)
and $3.11\ms$ for the $f=1$ pure DS in this case (orange dotted curve),
whereas intermediate mixed stars have lower $\mmax$.
Those are plotted in intervals of $f$ varying by 0.02.
One can classify the stars as either DM-core or DM-halo,
and for each fixed-$f$ curve a marker denotes the transition $R_D=R_N$
between these configurations, if present.
For example, for $f=0.1$ all stars are DM-core,
whereas for $f=0.2$ there are two transitions
at $(M,R)=(1.69\ms,10.64\km)$ and $(M,R)=(0.18\ms,15.69\km)$
from DM core to DM halo and back.
Many $(M,R_N)$ points are double- or triple-occupied,
as for example the $f=0.1,0.9,0.99994$ stars
with a common $(M,R_N) =(1.11\ms,11.07\km)$.
These multiple-occupied configurations,
characterized by different $f$,
belong to the class of the so-called ``twin" stars,
which differ substantially in their internal structure
and DM radius,
but would be indistinguishable by just mass and optical radius observations.
A more detailed discussion can be found in \cite{Liu23}.

Another interesting aspect are configurations of nearly pure NSs or DSs,
with a very small fraction of the minority component.
Those correspond to small amounts of minority matter
trapped in a container majority star,
and are characterized by small minority radii.
Those are shown in the figure by thin solid blue curves ($f\ra1$)
for DSs containing a small amount of pure NM,
plotted in intervals of $f=1-10^{-2,-3,...,-9}$,
or thin dotted red ($f\ra0$) curves which correspond to pure NM configurations
with a tiny amount of DM equal to $f=10^{-2,-3,...,-7}$.

According to this discussion,
one can prepare for a given $\mu$
a configuration plot in the $(M,R_N)$ plane \cite{Kain21,Liu23},
that shows all possible DM-core or DM-halo stars,
corresponding to the domain covered by solid curves in Fig.~\ref{f:mrnd},
for example.
This is done in Fig.~\ref{f:dns} for $\mu=0.8,1.6,3.2,6.4 \gev$.
According to the previous discussion,
magenta domains contain only DM-core stars,
delimited by the black solid curve ($f=0$),
whereas the green domains
(dark green $f<0.99$ and light green $f>0.99$ for clarity)
host only DM-halo stars.
In the cyan domain both kind of configurations are present.
One notes that for `large' $\mu$ nearly all stellar configurations
are DM core (magenta),
and added DM causes always a reduction of radius and maximum DNS mass.
However, these possible changes become smaller and smaller with increasing $\mu$.
With decreasing $\mu$ instead,
DM halo configurations (green) become increasingly dominant.
They feature a wide range of possible optical radii $R_N$
from zero (for very diluted NM, $f\ra1$)
to very large values $\gg R(f=0)$.
Also the possible maximum mass becomes larger than the pure NS one
for $\mu<1.11\gev$,
increasing with decreasing $\mu$.

Therefore, the possible DNS configurations depend strongly
on the DM particle mass $\mu$,
and this fact can be exploited to conversely deduce the values of $\mu$ and $f$
from eventual DNS observations.
This possibility will be examined in detail in the following.

\section{Results}
\label{Sec.3}

\subsection{Mass and radius}

We begin in Fig.~\ref{f:r14}
with the most significant observables mass and radius,
namely the maximum mass $\mmax$ (a)
and the optical radii $\r14$ (b) and $R^{(N)}_{2.0}$ (c)
as contour plots of the two model parameters $\mu$ and $f$.
According to the preceeding discussion,
and roughly dividing into the cases $\mu\gtrsim1\gev$ and $\mu\lesssim1\gev$,
in the first case $\mmax<\mmax(f=0)=2.36\ms$
(indicated by a red contour)
for any $f$
(mostly DM-core stars),
whereas in the second case $\mmax$ might reach very large values with
increasing $f$
(all DM-halo stars).
The (red) $M=1.4\ms$ contour identifies the maximum value of $\mu$
as a function of $f$,
for which a $1.4\ms$ DNS is possible.

We examine the optical radius $\r14$ of such stars in panel (b).
Again dividing roughly into the two cases,
for high masses $\mu\gtrsim1\gev$ there are only DM-core stars,
where the DM core pulls together also the nuclear matter,
always reducing the radius with increasing $f$
from the $f=0$ pure NS value $12.28\km$.
The maximum possible DM fraction $f$ is limited in this regime
by the condition that a stable $1.4\ms$ DNS can be formed.

On the other hand, most $\mu\lesssim1\gev$ configurations are DM-halo,
without upper limits on $f$,
and for sufficiently large $f$ also the optical radius is stretched out
to large values by the surrounding DM halo.
This concurs with very large $\mmax$ in panel (a).
These stars feature `low-density' NM
(described by the NS crust EOS in this work)
embedded in a larger DM `container' star.
Similar as pointed out for Fig.~\ref{f:mrnd},
in this domain for a fixed $\mu$ two or three configurations
with the same $\r14$ but different $f$ (and different $R_{1.4}$) exist;
therefore, only knowing $\mu$ and $\r14$,
a unique classification of a DNS is not possible here.

Qualitatively similar results for $R^{(N)}_{2.0}$ are shown in panel (c).
Naturally the parameter space is more restricted here
due to the $M=2\ms$ condition.

\subsection{Tidal deformability}

An observable closely related to the radius $R$ is the
tidal deformability (quadrupole polarizability) $\la$,
obtained from the tidal Love number $k_2$
\cite{Hartle67,Binnington09,Hinderer10,Damour10,Postnikov10},
\be
 \la = \frac23 \bigg(\frac{R}{M}\bigg)^5 k_2 \:,
\ee
where $R$ is the gravitational-mass radius, i.e.,
the outer DNS radius in our formalism.
The relevant equations to compute $k_2$
in the two-fluid formalism can be found in
\cite{Leung22,Karkevandi22,Dengler22,Zhang23},
for example.

In Fig.~\ref{f:l14}(a)
we display ${(k_2)}_{1.4}(f,\mu)$ as a contour plot,
just as $\r14$ in Fig.~\ref{f:r14}(b).
One can see that for not very large values of $f$
the effect of DM admixture is always a reduction of $k_2$.
Only for very dark stars with $f\gtrsim0.8$ values of
$k_2>0.086$ (the one of a pure NS)
are reached again.
However, $k_2$ is not directly observable,
but the relevant quantity is the tidal deformability
$\l14(f,\mu)$ shown in panel (b).
Obviously, the dependence of $\l14$
on the fifth power of the radius $R_{1.4}$ then
implies also a very strong dependence on $(f,\mu)$
that can clearly be seen in the figure.
For DM-core configurations $\l14$ is reduced
compared to the pure NS value 430
(red contour),
as is the radius $R_{1.4}=\r14$ reported in Fig.~\ref{f:r14}(b),
whereas for DM-halo configurations with large $f$ and
$R_{1.4}=R_{1.4}^{(D)}$,
$\l14$ becomes enormous.
Thus any non-tiny admixture of DM causes a substantial effect
that we analyze as follows.

A universal relation between the tidal deformability
and the compactness $M/R$ of pure NSs
was introduced in Ref.~\cite{Yagi13},
and in Ref.~\cite{Yagi17,Wei19} the following fit was proposed
\be
 \frac{M}{R} = 0.36 - 0.0355 \ln\la + 0.000705 (\ln\la)^2 \:,
\label{e:univ}
\ee
or equivalently
\be
 \log\la_\text{fit}(R) = 10.9 - 16.4 \sqrt{\frac{M}{R} + 0.087} \:,
\label{e:univ2}
\ee
which holds to within 7\% for a large set of NS nucleonic EOSs \cite{Yagi17}.
Recently an equivalent but more sophisticated fit
was derived in \cite{Godzieba21}.
Substantial deviations from this fit formula therefore indicate
`non-nucleonic' compact stars,
and we illustrate this by displaying in Fig.~\ref{f:l14}(c)
the ratio $\lr14$ between the DNS value
and the expected value of a pure NS with the same optical radius $\r14$
as the DNS.

In the $\mu\gtrsim1\gev$ DM-core regime
we have $\r14=R_{1.4}$
and therefore the reduction of $\l14$ simply reflects
the moderate reduction of the radius in this domain,
leading also to a moderate reduction of $\lr14$.
However, in the $\mu\lesssim1\gev$ DM-halo regime
with $\r14 \ll R_{1.4}$,
apart from the very large values of $\l14$ due to the extended halo,
the value of $\la_\text{fit}(\r14)$ is also too small
and this amplifies the ratio $\lr14$ even more,
to increases of several orders of magnitude
with $f\ra 1$ but in particular $\mu\ra0$.
Thus DNSs in this regime should exhibit very clear observational signatures
by just analyzing their optical radius and tidal deformability,
in the way just presented.

For a different illustration of this feature,
we compare in Fig.~\ref{f:lmr}
explicitly the universal relation $\la_\text{fit}(M/R_N)$
with the actual DNS relations for
either fixed $\mu=1\gev$ (a) or fixed $f=0.1$ (b),
varying the other variable.
The enormous enhancement of $\la$ relative to $\la_\text{fit}$
for DM-halo stars is evident in both cases,
increasing with $f\ra1$ and $\mu\ra0$, respectively,
as explained before.
The results for the $M=1.4\ms$ configurations
(square markers) in the figure
correspond to those analyzed before in Figs.~\ref{f:r14} and \ref{f:l14},
but the qualitative behavior is the same for any mass $M$
(or compactness $M/R$).
Equivalent results but limited to very small DM fractions
were also obtained recently in \cite{Routaray23,Thakur23}.
It is obvious that the GW170817 constraint
$\l14=70$--$580$ is compatible with a very wide range of parameters $(\mu,f)$
as can be seen more clearly in Fig.~\ref{f:l14}(a).

The figure in addition shows the relation between $\la$
and the true compactness $M/R$ (dashed curves),
should one be able to deduce the gravitational radius $R$ observationally.
In this case the actual values lie always below the fit formula
due to the large difference between $R$ and $R_N$ in DM-halo stars.
Thus also in this kind of analysis there would be a unique signature
of DM in DNSs.


\section{Summary}
\label{Sec.4}

We have analyzed the properties of DNSs with substantial DM fraction,
combining a well-constrained nucleonic EOS with a fermionic DM EOS
respecting an important constraint on the self-interaction cross section.
The analysis was focused on the most important and accessible observables:
mass, radius, and tidal deformability.
Already at this level,
there are very pronounced observational features
(violation of universal relations)
that would clearly discern a DNS from a `standard' NS,
provided the DM fraction is not tiny.
Registration of future GW events will allow to perform such analysis
with good precision,
allowing a search for DM in this way.

\vspace{10mm}
\section*{Acknowledgments}

This work is sponsored by
the National Key R\&D Program of China No.~2022YFA1602303 and
the National Natural Science Foundation of China under Grant
Nos.~11975077, 12147101, and 12205260.


\def\aap{A\&A}
\def\apjl{Astrophys. J. Lett.}
\def\araa{Annu. Rev. Astron. Astrophys.}
\def\epja{EPJA}
\def\epjc{EPJC}
\def\jcap{Journal of Cosmology and Astroparticle Physics}
\def\jcap{JCAP}
\def\mnras{MNRAS}
\def\npb{Nucl. Phys. B}
\def\physrep{Phys. Rep.}
\def\plb{Phys. Lett. B}
\def\ppnp{Prog. Part. Nucl. Phys.}
\def\rpp{Rep. Prog. Phys.}
\bibliographystyle{apsrev4-1}
\bibliography{dmf}

\end{document}